\documentclass[         %
aps,                    %  American Physical Society
jpa,                    %  Journal Physics A
showpacs,               %  Displays PACS after abstract
nofootinbib,            %  Does not treat footnotes as references
showkeys,               %
preprintnumbers,        %
floatfix]               %  Fixes float errors
{revtex4}               %  REVTEX 4 Package used
\tighten
\draft
\usepackage{graphicx} 

\begin{document}

\title{Character Expansions in Physics}


\author{A.B. Balantekin}
\address{Physics Department, University of Wisconsin, Madison WI 53706 USA}

\begin{abstract}
Expanding products of invariant functions of a group element as a series in the basis of characters of the irreducible representations of a group is widely used in many areas of physics and related fields. In this contribution a formula to generate such expansions and its various applications are briefly reviewed.  
\end{abstract}

\maketitle

%%%%%%%%%%%%%%%%%%%%%%%%%%%%%%%%%%%%%%%%%%%%
%% MAINMATTER
%%%%%%%%%%%%%%%%%%%%%%%%%%%%%%%%%%%%%%%%%%%%

\section{Introduction}
The problem of integrating functions of group elements over the entire group manifold shows up 
in many branches of physics in attempts as diverse as exploring various aspects of QCD and 
other gauge theories 
\cite{Balantekin:2000vn,Akemann:2004nw,Budczies:2001tb,Lenaghan:2001ur,Akemann:2000df,Akemann:2001ir}, 
quantum transport in stochastic cavities \cite{Khor},
multiple input and multiple output systems in communications \cite{mimo}, 
statistical theory of nuclear reactions \cite{mello},
various aspects of statistical physics \cite{Balantekin:2001af,Splittorff:2002eb,Simon:2004ma}, 
matrix models \cite{Morozov:2009jv,Fyodorov:2006pk}, 
lattice gauge theory \cite{Bars:1980yy,Bars:1979xb}, 
functional integral approaches to gauge theories \cite{Dashen:1975hd}, 
random matrix models \cite{Guhr:1997ve,Mitchell:2010um}, 
as well as in astronomy \cite{Desjacques:2007zg} and 
in the study of cognitive radios \cite{cr}. In the following, one approach to calculating such integrals is presented. 

Characters are defined as the traces of the representation matrices. 
For example representations of the U(N) group are 
labeled by a partition into $N$ parts: ($n_1,n_2,\cdots,n_N$) where 
$n_1 \ge n_2 \ge \cdots \ge n_N$ are visualized as the number of rows in Young
Tableaux. The character 
of the irreducible representation
corresponding to the partition ($n_1,n_2,\cdots,n_N$) of non-negative
integers is given by Weyl's formula:
\begin{equation}
\label{weylsfor1}
\chi_{(n_1,n_2,\cdots,n_N)} (U) = \frac{\det ( t_i^{n_j+N-j})}{\Delta 
(t_1, \cdots, t_N)},
\end{equation}
where $t_i,i=1,\cdots,N$, are the eigenvalues of the group element $U$
in the fundamental representation and the quantity $\Delta (t_1, \cdots, t_N)$ is
the Vandermonde determinant in the arguments $t_1, \cdots, t_N$: 
\begin{equation}
\label{vandet}
\Delta (t_1, \cdots, t_N) = \det ( t_i^{N-j}).
\end{equation}
In these equations the arguments of the determinants indicate the
$(ij)$-th element of the matrix the determinant of which is
calculated. An alternative form for the character formula is given by
\begin{equation}
\label{weylsfor2}
\chi_{(n_1,n_2,\cdots,n_N)} (U) = \det ( h_{n_j+i-j}) ,
\end{equation}
where $h_n$ is the complete symmetric function in the arguments $t_1,
\cdots, t_N$ of degree $n$. (The precise definition of the complete symmetric functions is given in the last section below). 

In most applications the quantity that is integrated over the group manifold can be written 
as products of functions that are invariant
under the group transformations, e.g.:
\begin{equation}
\label{inva}
\int dU f(U^{\dagger}) g(U). 
\end{equation}
Sometimes these functions can be expanded in terms of the characters of the group:
\begin{equation}
\label{sum}
f (U) = \sum_r f_r \chi_r (U), 
\end{equation}
where the sum is over all irreducible representations, $r$. In such cases group integration can be easily carried out using the orthogonality of the characters:
\begin{equation}
\int dU \chi_r^* (U) \chi_s (U) = \delta_{rs} .
\end{equation}
The question addressed here is how to determine the expansion coefficients in Eq. (\ref{sum}). 

\section{A Character expansion formula}

Let us, for example, consider the quantity $\exp (x {\rm Tr} U)$. The usual Taylor expansion of the exponential includes terms with $({\rm Tr} U)^n$. For the group U(N),  $({\rm Tr} U)^n$ can be written as a sum of the characters of all the U(N) representations satisfying the condition $n_1+n_2+\cdots+n_N=n$ (n-boxes in 
the corresponding Young tableaux) \cite{mello,Bars:1980yy} and the ordinary Taylor expansion can be considered a character expansion. One can use similar tricks for writing down other character expansions, however such a procedure quickly becomes too tedious and one may ask if there is a better approach. 
Indeed such an approach exits and starts with the power series expansion
\begin{equation}
\label{powerser}
G(x,t) = \sum_n A_n(x) t^n ,
\end{equation}
where the range of $n$ in the sum is not yet specified. In Eq. (\ref{powerser}), $x$ stands for
all the parameters needed to specify the coefficients $A_n$. We assume
that this series is convergent for $|t|=1$. After some manipulations one obtains the following character expansion formula \cite{Balantekin:2000vn}: 
\begin{equation}
\label{7}
\left( \prod_{i=1}^N G(x,t_i) \right) = \sum_{m_1=0} \sum_{m_2 =0}
\cdots \sum_{m_{N-1}=0} \sum_{n_N} \det (A_{n_j +i -j}) \left(\det U
\right)^{n_N} \chi_{(\ell_1, \ell_2, \cdots, \ell_N)} (U). 
\end{equation}
In Eq. (\ref{7}), $t_i, i=1,\cdots,N$ are the eigenvalues of the fundamental representation $U$, 
the integers $m_i, i=1,\cdots,N-1$ are all non-negative, 
\begin{equation}
\label{8a}
n_i = m_i + m_{i+1} + \cdots + m_{N-1} + n_N, 
\end{equation}
and the integers
\begin{eqnarray}
\label{8}
\ell_i &=& \sum_{j=i}^{N-1} m_j, \>\>  \> i=1,\cdots,N-1,   \\
\ell_N &=& 0  
\end{eqnarray}
label the irreducible representations of U(N). If, in addition, all the $A_n$ in Eq. (\ref{powerser}) are non-negative, then Eq. (\ref{7}) takes a particularly simple form \cite{Balantekin:1983km}:
\begin{equation}
\label{9}
\left( \prod_{i=1}^N G(x,t_i) \right) = \sum_{n_1=0} \sum_{n_2 =0}
\cdots \sum_{n_N=0} \det (A_{n_j +i -j}) \chi_{(n_1,n_2,\cdots,n_N)}
(U) .
\end{equation} 
It is also possible to generalize Eqs. (\ref{7}) and (\ref{9}) to orthogonal and symplectic groups 
\cite{Balantekin:2001id}. 

The characters of the covariant class I
representations of the supergroup U(N/M) are given by a formula similar to
Eq. (\ref{weylsfor2}) except that the complete symmetric functions are
replaced by the graded homogeneous symmetric functions
\cite{Baha Balantekin:1980qy}. 
The complete symmetric functions can be written in terms of the traces of the
fundamental representation. The graded homogenous symmetric functions are given by similar
expressions except that traces are replaced by supertraces
\cite{Baha Balantekin:1980qy,Baha Balantekin:1980pp}. 
Since the character expansion formula above is  
basically combinatorial in nature, it is also applicable in principle to
the covariant representations of the supergroup U(N/M). 
For recent work using character expansions for supergroups, see, example, 
Ref. \cite{Lehner:2008fp}.  Physics literature also contains many recent applications of the invariant integration over groups and supergroups (see, for example, Refs. 
\cite{Alfaro:1994ca}, \cite{Wei:2004cc}, \cite{Aubert:2004ra}, \cite{Schlittgen:2002tj}, 
and \cite{ZinnJustin:2002pk}).

\section{Examples}

The complete homogeneous symmetric function, $h_n (x)$, of degree $n$
in the arguments $x_i, i=1, \cdots, N$, is defined as the sum of the
products of the variables $x_i$, taking $n$ of them at a time. Its generating function is 
\begin{equation}
\label{10} 
\frac{1}{\prod_{i=1}^N (1 - x_i z)} = \sum_n h_n(x) z^n .
\end{equation}
Taking $x_i$ to be eigenvalues of the fundamental representation $U$ of the group SU(N), 
and using Eq. (\ref{9}), Eq. (\ref{10}) can be written as a character expansion formula: 
\begin{equation}
\label{11} 
\frac{1}{\det (1-zU)} = \sum_n \chi_{(n,0,0,\cdots)} z^n .
\end{equation}
What if we need (for example in a non-linear theory) to include higher powers of $U$, i.e. we want to expand the quantity $\det (1- 2x U +  U^2)$, or its inverse, in terms of characters? Taking 
$G(x,t)$ of Eq. (\ref{powerser})  to be $1 -2xt + t^2$ (i.e. $A_0=1$, $A_1 = -2x$, $A_2 =1$, and 
$A_n = 0$ for $n \ge 3$), one can immediately write a character expansion for 
$\det (1- 2x U +  U^2)$. Although such an expansion may look complicated, further inspection reveals that characters corresponding to Young tableaux with more than two boxes at each row do not appear in it.  For the inverse quantity, one can start with the generating function for Chebyshev polynomials of the second 
kind, $u_n(x)$: 
\begin{equation}
\label{12} 
\frac{1}{1-2tx+x^2} = \sum_{n=0}^{\infty} u_n(x) t^n. 
\end{equation}
From Eq. (\ref{9}), one then gets 
\begin{equation}
\label{13}
 \frac{1}{\det(1-2xU+U^2)} = \sum_{n_1=0} \sum_{n_2 =0}
\cdots \sum_{n_N=0} \det (u_{n_j +i -j}) \chi_{(n_1,n_2,\cdots,n_N)}
(U) .
\end{equation} 
This expression is again simpler than it looks. For example characters corresponding to single column 
Young tableaux with more than two boxes do not appear. (This can be proved by noting that the quantity $1 -2xt + t^2$ and the generating function of the Chebyshev polynomials of the second kind are inverses of each other and, as such, there are relations between coefficients of $t^n$ in each expansion. These relationships can be expressed as determinants). Further examples of character expansions in physics applications are given in Ref. \cite{Balantekin:2000vn}

%%%%%%%%%%%%%%%%%%%%%%%%%%%%%%%%%%%%%%%%%%%%%%%%
%% BACKMATTER
%%%%%%%%%%%%%%%%%%%%%%%%%%%%%%%%%%%%%%%%%%%%%%%%

\section*{Acknowledgments}
This work was supported in part
by the U.S. National Science Foundation Grant No. PHY-0855082
and
in part by the University of Wisconsin Research Committee with funds
granted by the Wisconsin Alumni Research Foundation.

%%%%%%%%%%%%%%%%%%%%%%%%%%%%%%%%%%%%%%%%%%%
%% The following lines show an example how to produce a bibliography
%% without the help of the BibTeX program. This could be used instead
%% of the above.
%%%%%%%%%%%%%%%%%%%%%%%%%%%%%%%%%%%%%%%%%%%


\begin{thebibliography}{99}

%\cite{Balantekin:2000vn}
\bibitem{Balantekin:2000vn}
  A.~B.~Balantekin,
  %``Character expansions, Itzykson-Zuber integrals, and the QCD partition
  %function,''
  \emph{Phys.\ Rev.\  D} \textbf{62}, 085017 (2000)
  [arXiv:hep-th/0007161].
  %%CITATION = PHRVA,D62,085017;%%

%\cite{Akemann:2004nw}
\bibitem{Akemann:2004nw}
  G.~Akemann, Y.~V.~Fyodorov and G.~Vernizzi,
  %``On matrix model partition functions for QCD with chemical potential,''
  \emph{Nucl.\ Phys.\  B} \textbf{694}, 59 (2004)
  [arXiv:hep-th/0404063].
  %%CITATION = NUPHA,B694,59;%%

%\cite{Budczies:2001tb}
\bibitem{Budczies:2001tb}
  J.~Budczies, S.~Nonnenmacher, Y.~Shnir and M.~R.~Zirnbauer,
  %``(1+1)-dimensional baryons from the SU(N) color-flavor transformation,''
  \emph{Nucl.\ Phys.\  B} \textbf{635}, 309 (2002)
  [arXiv:hep-lat/0112018].
  %%CITATION = NUPHA,B635,309;%%
  
  %\cite{Lenaghan:2001ur}
\bibitem{Lenaghan:2001ur}
  J.~Lenaghan and T.~Wilke,
  %``Mesoscopic QCD and the Theta vacua,''
 \emph{ Nucl.\ Phys.\  B} \textbf{624}, 253 (2002)
  [arXiv:hep-th/0108166].
  %%CITATION = NUPHA,B624,253;%%

%\cite{Akemann:2000df}
\bibitem{Akemann:2000df}
  G.~Akemann, D.~Dalmazi, P.~H.~Damgaard and J.~J.~M.~Verbaarschot,
  %``QCD(3) and the replica method,''
  \emph{Nucl.\ Phys.\  B} \textbf{601}, 77 (2001)
  [arXiv:hep-th/0011072].
  %%CITATION = NUPHA,B601,77;%%
  
%\cite{Akemann:2001ir}
\bibitem{Akemann:2001ir}
  G.~Akemann, J.~T.~Lenaghan and K.~Splittorff,
  %``Dashen's phenomenon in gauge theories with spontaneously broken chiral
  %symmetries,''
 \emph{Phys.\ Rev.\  D} \textbf{65}, 085015 (2002)
  [arXiv:hep-th/0110157].
  %%CITATION = PHRVA,D65,085015;%%
  
\bibitem{Khor}
B.A. Khoruzhenko, D.V. Savin, and H.-J. Sommers, 
\emph{Phys. Rev. B} \textbf{80}, 125301 (2009). 

\bibitem{mimo}
S.H. Simon, A.L. Moustakas, L. Marinelli, 
\emph{IEEE Tran. Information Theo.} \textbf{52},   5336 (2006); 
M.R. Mckay {\it et al.}, 
\emph{IEEE Tran. Vehicular Tech.} \textbf{56}, 2555 (2007); 
L.G. Ordonez, D.P. Palomar, and J.R. Fonollosa, 
\emph{IEEE Tran. Signal Process.} \textbf{57}, 672 (2009). 

\bibitem{mello}
 M. Gaudin  and   P. A. Mello, \emph{J. Phys. G: Nucl. Phys.}\textbf{7}, 1085 (1981).

%\cite{Balantekin:2001af}
\bibitem{Balantekin:2001af}
  A.~B.~Balantekin,
  %``Partition Functions in Statistical Mechanics, Symmetric Functions, and
  %Group Representations,''
 \emph{Phys.\ Rev.\  E} \textbf{64}, 066105 (2001)
  [arXiv:cond-mat/0109112].
  %%CITATION = PHRVA,E64,066105;%%

%\cite{Splittorff:2002eb}
\bibitem{Splittorff:2002eb}
  K.~Splittorff and J.~J.~M.~Verbaarschot,
  %``Replica limit of the Toda lattice equation,''
 \emph{ Phys.\ Rev.\ Lett.}\textbf{90}, 041601 (2003)
  [arXiv:cond-mat/0209594].
  %%CITATION = PRLTA,90,041601;%%

%\cite{Simon:2004ma}
\bibitem{Simon:2004ma}
  S.~H.~Simon and A.~L.~Moustakas,
  %``Eigenvalue Density of Correlated Complex Random Wishart Matrices,''
  \emph{Phys.\ Rev.\  E} \textbf{69}, 065101 (2004)
  [arXiv:math-ph/0401038].
  %%CITATION = PHRVA,E69,065101;%%


%\cite{Morozov:2009jv}
\bibitem{Morozov:2009jv}
  A.~Y.~A.~Morozov,
  %``Unitary Integrals and Related Matrix Models,''
  \emph{Theor.\ Math.\ Phys.} \textbf{162}, 1 (2010)
  [\emph{Teor.\ Mat.\ Fiz.} \textbf{161}, 3 (2010)]
  [arXiv:0906.3518 [hep-th]].
  %%CITATION = TMFZA,161,3;%%

%\cite{Fyodorov:2006pk}
\bibitem{Fyodorov:2006pk}
  Y.~V.~Fyodorov and B.~A.~Khoruzhenko,
  %``A few remarks on colour-flavour transformations, truncations of random
  %unitary matrices, Berezin reproducing kernels and Selberg type integrals,''
  \emph{J.\ Phys.\ A} \textbf{40}, 669 (2007)
  [arXiv:math-ph/0610045].
  %%CITATION = JPAGB,A40,669;%%


%\cite{Bars:1980yy}
\bibitem{Bars:1980yy}
  I.~Bars,
  %``U(N) Integral For Generating Functional In Lattice Gauge Theory,''
  \emph{J.\ Math.\ Phys.} \textbf{21}, 2678 (1980).
  %%CITATION = JMAPA,21,2678;%%
  
%\cite{Bars:1979xb}
\bibitem{Bars:1979xb}
  I.~Bars and F.~Green,
  %``Complete Integration Of U (N) Lattice Gauge Theory In A Large N Limit,''
  \emph{Phys.\ Rev.\  D}  \textbf{20}, 3311 (1979).
  %%CITATION = PHRVA,D20,3311;%%
  
%\cite{Dashen:1975hd}
\bibitem{Dashen:1975hd}
  R.~F.~Dashen, B.~Hasslacher and A.~Neveu,
  %``The Particle Spectrum In Model Field Theories From Semiclassical Functional
  %Integral Techniques,''
  \emph{Phys.\ Rev.\  D} \textbf{11}, 3424 (1975).
  %%CITATION = PHRVA,D11,3424;%%


%\cite{Guhr:1997ve}
\bibitem{Guhr:1997ve}
  T.~Guhr, A.~Muller-Groeling and H.~A.~Weidenmuller,
  %``Random matrix theories in quantum physics: Common concepts,''
  \emph{Phys.\ Rept.} \textbf{299}, 189 (1998)
  [arXiv:cond-mat/9707301].
  %%CITATION = PRPLC,299,189;%%

%\cite{Mitchell:2010um}
\bibitem{Mitchell:2010um}
  G.~E.~Mitchell, A.~Richter and H.~A.~Weidenmueller, 
 \emph{Rev. Mod. Phys.} \textbf{82}, 2845 (2010).  
  %``Random Matrices and Chaos in Nuclear Physics: Nuclear Reactions,''
  arXiv:1001.2422 [nucl-th].
  %%CITATION = ARXIV:1001.2422;%%

%\cite{Desjacques:2007zg}
\bibitem{Desjacques:2007zg}
  V.~Desjacques,
  %``Environmental dependence in the ellipsoidal collapse model,''
  \emph{Mon.\ Not.\ Roy.\ Astron.\ Soc.} \textbf{388}, 638 (2008)
  [arXiv:0707.4670 [astro-ph]].
  %%CITATION = MNRAA,388,638;%%

\bibitem{cr}
R. Couillet and M. Debbah, 
\url{http://hal-supelec.archives-ouvertes.fr/hal-00448807/en/}. 

%\cite{Balantekin:1983km}
\bibitem{Balantekin:1983km}
  A.~B.~Balantekin,
  %``Character Expansion For U(N) Groups And U(N/M) Supergroups,''
  \emph{J.\ Math.\ Phys.} \textbf{25}, 2028--2030 (1984).
  %%CITATION = JMAPA,25,2028;%%

%\cite{Balantekin:2001id}
\bibitem{Balantekin:2001id}
  A.~B.~Balantekin and P.~Cassak,
  %``Character expansions for the orthogonal and symplectic groups,''
 \emph{J.\ Math.\ Phys.}  \textbf{43}, 604--620 (2002)
  [arXiv:hep-th/0108130].
  %%CITATION = JMAPA,43,604;%%

%\cite{Baha Balantekin:1980qy}
\bibitem{Baha Balantekin:1980qy}
  A. B.  Balantekin and I.~Bars,
  %``Dimension And Character Formulas For Lie Supergroups,''
  J.\ Math.\ Phys.\  {\bf 22}, 1149 (1981).
  %%CITATION = JMAPA,22,1149;%%

%\cite{Baha Balantekin:1980pp}
\bibitem{Baha Balantekin:1980pp}
  A. B.  Balantekin and I.~Bars,
  %``Representations Of Supergroups,''
  J.\ Math.\ Phys.\  {\bf 22}, 1810 (1981).
  %%CITATION = JMAPA,22,1810;%%

%\cite{Lehner:2008fp}
\bibitem{Lehner:2008fp}
  C.~Lehner, T.~Wettig, T.~Guhr and Y.~Wei,
  %``Character expansion method for supergroups and extended superversions of
  %the Leutwyler-Smilga and Berezin-Karpelevich integrals,''
  \emph{J.\ Math.\ Phys.} \textbf{49}, 063510 (2008)
  [arXiv:0801.1226 [math-ph]].
  %%CITATION = JMAPA,49,063510;%%

%\cite{Alfaro:1994ca}
\bibitem{Alfaro:1994ca}
  J.~Alfaro, R.~Medina and L.~F.~Urrutia,
  %``The Itzykson-Zuber integral for U(m/n),''
  \emph{J.\ Math.\ Phys.} \textbf{36}, 3085 (1995)
  [arXiv:hep-th/9412012].
  %%CITATION = JMAPA,36,3085;%%

%\cite{Wei:2004cc}
\bibitem{Wei:2004cc}
  Y.~Wei and T.~Wettig,
  %``Bosonic color-flavor transformation for the special unitary group,''
  \emph{J.\ Math.\ Phys.} \textbf{46}, 072306 (2005)
  [arXiv:hep-lat/0411038].
  %%CITATION = JMAPA,46,072306;%%

%\cite{Aubert:2004ra}
\bibitem{Aubert:2004ra}
  S.~Aubert and C.~S.~Lam,
  %``Invariant and Group Theoretical Integrations over the U(n) Group,''
  \emph{J.\ Math.\ Phys.} \textbf{45}, 3019 (2004)
  [arXiv:math-ph/0405036].
  %%CITATION = JMAPA,45,3019;%%




  %\cite{Schlittgen:2002tj}
      \bibitem{Schlittgen:2002tj}
        B.~Schlittgen and T.~Wettig,
        %``Generalizations of some integrals over the unitary group,''
        \emph{J.\ Phys.\ A} \textbf{36}, 3195 (2003)
        [arXiv:math-ph/0209030].
        %%CITATION = JPAGB,A36,3195;%%


      %\cite{ZinnJustin:2002pk}
      \bibitem{ZinnJustin:2002pk}
        P.~Zinn-Justin and J.~B.~Zuber,
        %``On some integrals over the U(N) unitary group and their large N limit,''
        \emph{J.\ Phys.\ A} \textbf{36}, 3173 (2003)
        [arXiv:math-ph/0209019].
        %%CITATION = JPAGB,A36,3173;%%


\end{thebibliography}
\end{document}